\documentclass[prl,aps,superscriptaddress,twocolumn,color]{revtex4-2}

\usepackage{amsmath}
\usepackage{amsfonts}
\usepackage{bm}
\usepackage{graphicx}
\usepackage{array,color}
\usepackage{braket}
\usepackage[table]{xcolor}
\usepackage{xcolor}
\usepackage{chemformula}

\begin{document}
\title{Engineering mixing properties of fluids by spatial modulations}

\author{Abid Ali}
\affiliation{Department of Engineering Science, University of
Electro-Communications, Tokyo 182-8585, Japan}

\author{Hiroki Saito}
\affiliation{Department of Engineering Science, University of
Electro-Communications, Tokyo 182-8585, Japan}

\date{\today}

\begin{abstract}
We propose a method to change the effective interaction between two
fluids by modulation of their local density distributions with
external periodic potentials, whereby the mixing properties can be
controlled.
This method is applied to a mixture of dilute bosonic gases, and
binodal and spinodal curves emerge in the phase diagram.
Spinodal decomposition into a mixed-bubble state becomes possible, in
which one of the coexisting phases has a finite mixing ratio.
A metastable mixture is also realized, which undergoes phase
separation via nucleation.
\end{abstract}

\maketitle

A binary mixture becomes thermodynamically unstable and separates into
two stable phases, when a control parameter, such as temperature, is
quenched across the critical point.
Such spontaneous phase separation of mixtures is known as spinodal
decomposition \cite{1,2,3,4}.
On the other hand, if the binary mixture is prepared in a metastable
state, finite perturbation is required for nucleation and growth to
proceed to phase separation \cite{5,6,7}.
Spinodal decomposition and nucleation are the two major mechanisms
responsible for phase separation in multicomponent systems.
The boundary between the spinodal and nucleation regions in the phase
diagram is called a spinodal curve, and that between the nucleation
and stable regions is called a binodal curve.
These three regions appear when the free energy has both concave and
convex shapes as a function of the mixing ratio.

The separated and mixed fluids have different free energies, since the
energy and entropy are changed by mixing.
Here we focus on the mixing energy, which is the energy difference
between separated and mixed fluids.
The mixing energy is determined by the interaction between constituent
particles of two fluids and, which is generally difficult to control.
The purpose of this Letter is to alter the mixing energy by a simple
method --- modulation of the local density distributions by external
periodic potentials, whereby mixing properties of fluids are
controlled.
Let us consider a mixing energy
$E_{\rm mix}(n_1(\bm{r}), n_2(\bm{r}))$,
which is dependent on the density distributions $n_1(\bm{r})$ and
$n_2(\bm{r})$ of components 1 and 2.
If we apply external potentials that locally modulate the density
distributions, the local overlap between $n_1(\bm{r})$ and
$n_2(\bm{r})$ is modulated.
On a scale much larger than the modulation wavelength, the effective
mixing energy is given by the spatial average
$\langle E_{\rm mix}(n_1(\bm{r}), n_2(\bm{r})) \rangle_{\bm{r}}$.
Therefore, the global mixing energy can be changed, which alters the
global mixing properties.

This method is applied to a binary mixture of Bose-Einstein
condensates (BECs) of ultracold gases~\cite{Myatt, Hall, Modugno,
  9,10,11,12,13}, which is a clean and highly controllable system.
In this system, external periodic potentials can be easily generated
and precisely controlled using optical lattices~\cite{Catani},
making the system suitable for the present purpose.
In a mixture of dilute BECs, in which simple mean-field theory is
applicable, the mixing property for a homogeneous system is trivial,
i.e., the mixture is either stable or unstable against phase
separation, irrespective of the mixing ratio, and there are no binodal
and spinodal curves in the phase diagram (with respect to the mixing
ratio).
It was recently predicted that a beyond-mean-field
effect~\cite{Petrov} can modify the energy curve and an interesting
separated phase (mixed-bubble state) is possible~\cite{15,16}, in
which one of coexisting phases has a finite mixing ratio.
However, the beyond-mean-field effect is a higher-order effect of the
density and emerges in the high-density regime.
Experimentally, atomic loss due to three-body recombination is
significant for such a density, especially near Feshbach resonance,
which limits the lifetime of the system \citep{17,18,19}.

Here we show that binodal and spinodal physics emerge in a binary
mixture of dilute BECs simply by application of a component-dependent
periodic external potential, which modulates the mixing energy
depending on the mixing ratio.
As a result, spinodal decomposition into the mixed-bubble state
becomes possible without beyond-mean-field effects.
Moreover, the mixture can be brought to a metastable state, in which a
finite perturbation is required to cross the energy barrier against
phase separation, which leads to nucleation and growth toward the
separated phase.

\begin{figure}[tb]
\includegraphics[width=7.5cm]{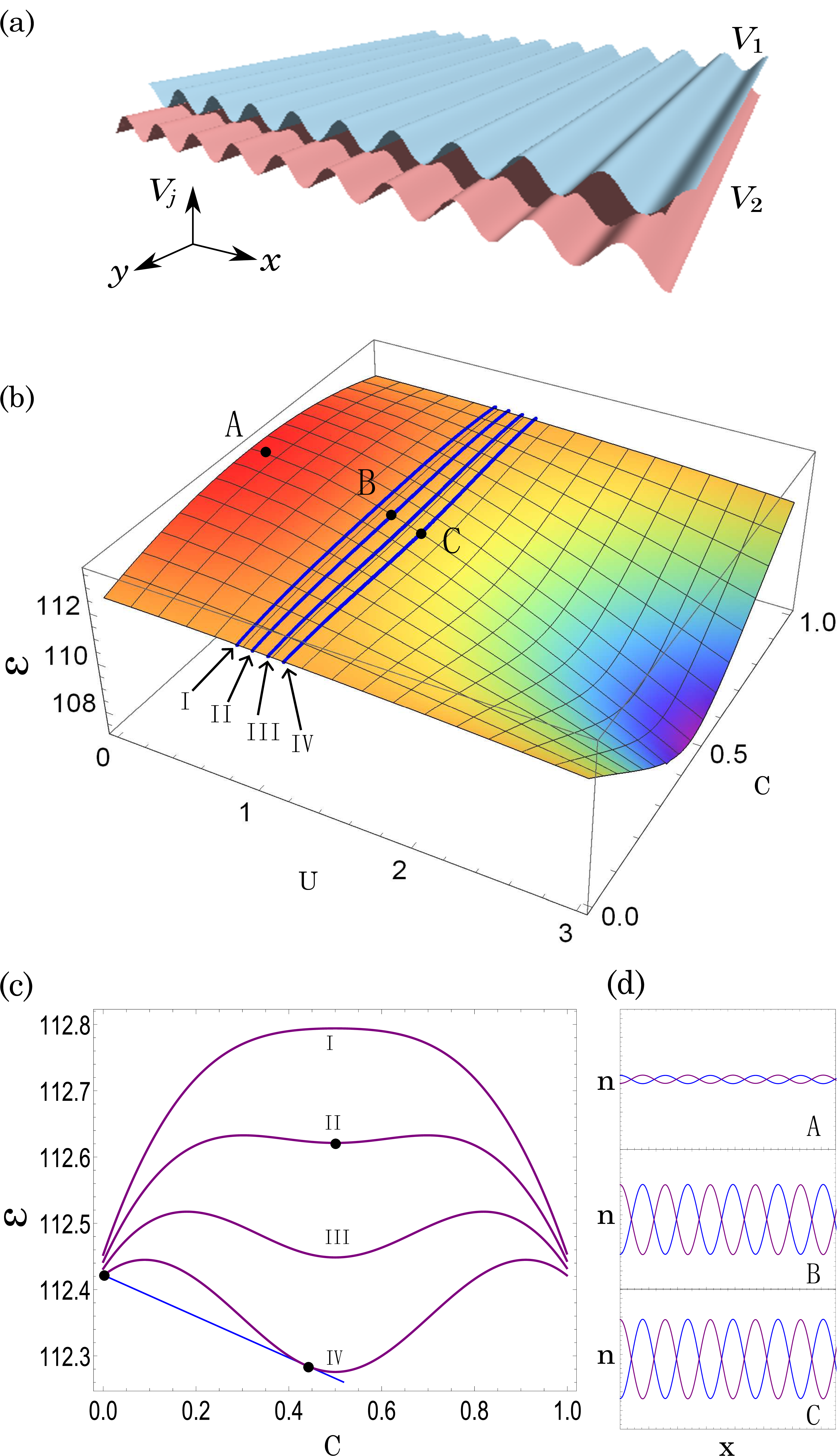} 
\caption{ (a) Schematic illustration of external periodic potentials
  $V_1 = U \cos^2(k x)$ for component 1 and $V_2 = -U \cos^2(k x)$ for
  component 2.
  (b) Variational energy $\varepsilon$ in Eq.~(\ref{varene}) as a
  function of potential strength $U$ and composition $C$, where
  variational parameters $a_1$ and $a_2$ are selected to minimize
  $\varepsilon$.
  The curves I, II, III, and IV correspond to $U = 0.9$, $U = 1$, $U =
  1.1$, and $U = 1.2$, respectively, which are shown in (c) as
  a function of $C$.
  The density distributions $n_1$ and $n_2$ are shown in (d), where
  panels A, B, and C correspond to the points marked in (b).
  The tangential line and circles on curves II and IV in (c)
  correspond to the mixed-bubble state and metastable state presented
  in Figs.~\ref{fig:bubble} and \ref{fig:ncl}, respectively.
  The interaction parameters are $g_{11} = g_{22} \equiv g = 15$ and
  $g_{12} = 15.3$.
\label{fig:fig1}}
\end{figure}

We consider a binary mixture of dilute Bose gases at zero temperature,
which can be described by the macroscopic wave functions $\Psi_{1}$
and $\Psi_{2}$ in the mean-field approximation.
The energy of the system can be written as~\cite{21,22}
\begin{eqnarray}
E & = & \int d\bm{r} \Biggl[ \sum_{j=1}^2 \Biggl(
  -\frac{\hbar^{2}}{2m} \Psi_{j}^{*}\nabla^{2}\Psi_{j}
  + V_{j}\vert\Psi_{j}\vert^{2}
  + \dfrac{g_{jj}}{2} \vert\Psi_{j}\vert^{4} \Biggr)
\nonumber \\ 
& & + g_{12}\vert\Psi_1 \vert^{2}\vert\Psi_2 \vert^{2}\Biggr],
\label{eq:one}
\end{eqnarray}
where $m$ is the atomic mass.
The wave function $\Psi_{j}(\bm{r}, t)$ satisfies $\int d\bm{r}
|\Psi_j(\bm{r}, t)|^2 = N_j$, where $N_j$ is the number of atoms for
component $j$.
The interaction coefficients are defined as $g_{jj'}=4\pi \hbar^{2}
a_{jj'} / m$, where $a_{jj'}$ is the $s$-wave scattering length
between components $j$ and $j'$.
In the present work, we consider the repulsive atomic interactions
with positive scattering lengths, $g_{ij} > 0$.
We consider a situation in which a periodic potential
$V_{j}(\bm{r}) = U_{j} \cos^{2}(kx)$ is applied to the system, where
$U_1 > 0$ and $U_2 < 0$ (see Fig.~\ref{fig:fig1}(a)).
Such a component-dependent periodic potential can be produced by
laser beams with a selected wave number $k$ and
polarization~\cite{23,24, Gadway, 25,26,27,Kim20,Kim}.

For a homogeneous system without an optical lattice, the energy
density is given by $\varepsilon = (g_{11} n_1^2 + g_{22} n_2^2)
/2+g_{12} n_1 n_2$, which has a constant curvature $g_{11} g_{22} -
g_{12}^2$ with respect to the uniform densities $n_1$ and $n_2$.
Therefore, there are only two ways to minimize $\omega \equiv
\varepsilon - \mu_1 n_1 - \mu_2 n_2$ for chemical potentials $\mu_j$.
For a positive curvature, there can be a single minimum with nonzero
$n_1$ and $n_2$, which corresponds to the uniformly mixed state.
For a negative curvature and appropriate $\mu_j$, $\omega$ can be
simultaneously minimized at the two points $n_1 = 0$ (with $n_2 \neq
0$) and $n_2 = 0$ (with $n_1 \neq 0$), which corresponds to the
separated state.
Thus, the phase diagram is trivial: the ground state is either the
uniformly mixed state or the totally separated state, which is
determined only by the sign of the curvature $g_{11} g_{22} -
g_{12}^2$, and is independent of the mixing ratio $N_1 / N_2$.
In the presence of component-dependent periodic external potentials,
as shown in Fig.~\ref{fig:fig1}(a), the density distributions
$n_1(\bm{r})$ and $n_2(\bm{r})$ are modulated (see
Fig.~\ref{fig:fig1}(d)), and their local overlap can be decreased.
As a result, the global mixing energy of the system is changed, and
the phase diagram will be altered.

To present this clearly, we employ a simple variational method.
The wave function is approximated as
\begin{equation} \label{psivar}
  \Psi_{j}(\bm{r})= \sqrt{n_{j} [1+a_{j}\cos(2kx)]},
\end{equation}
where $a_{j}$ is the real variational parameter satisfying $\vert
a_{j}\vert < 1$.
Substituting the variational wave function $\Psi_{j}(\bm{r})$ into the
energy of the system in Eq.~(\ref{eq:one}) gives
\begin{equation}
\begin{aligned}
 \varepsilon \equiv \frac{E}{V} ={} &
 \sum_{j=1}^2\Big[\frac{n_{j}}{2}\big(1-\sqrt{1-a_{j}^2}\big)+\frac{U_{j}}{4}n_{j}a_{j}\Big]\\
          &+\frac{g}{2}\Big[n_{1}^2\big(1+\frac{a_{1}^2}{2}\big)+n_{2}^2\big(1+\frac{a_{2}^2}{2}\big)\\
          &+g_{12}n_{1}n_{2}\big(1+\frac{a_{1}a_{2}}{2}\big)\Big],
\end{aligned}
\label{varene}
\end{equation} 
where $g_{11} = g_{22} \equiv g$ is assumed, $V$ is the volume of the
system, and the length, energy, and density are normalized by
$k^{-1}$, $\hbar^2 k^2 / m$, and $(N_1 + N_2) / V$, respectively.
It is also assumed that $|g_{12} - g| \ll g$, which maintains the
total density $|\Psi_1|^2 + |\Psi_2|^2$ to be almost uniform.
In this case, the densities can be parametrized by the composition $C$
as $n_1 = C$ and $n_2 = 1 - C$ with $0 \leq C \leq 1$.
In the following, we also assume $U_1 = -U_2 \equiv U$.
These assumptions are only to reduce the number of parameters and are
not crucial for the main results obtained later.

Figures \ref{fig:fig1}(b) and \ref{fig:fig1}(c) show the variational
energy $\varepsilon$ as a function of the potential strength $U$ and
composition $C$, where $\varepsilon$ has been minimized with respect
to the variational parameters $a_j$.
For a small value of $U$, the energy curve is concave (curve I), since
the interaction parameters satisfy $g_{11} g_{22} - g_{12}^2 < 0$.
Therefore, $C = 0$ and 1 minimize the energy, which indicates that
the mixture is energetically unstable against phase separation.
As the value of $U$ is increased, the energy curve is modified, and
the metastable state appears for $U=1$ (curve II).
For $U \gtrsim 1.1$, the energy around $C = 0.5$ decreases and goes
below those for $C = 0$ and 1 (curves III and IV).
It should be noted that such a concave-convex shape of $\varepsilon$
is due to the nonlinear dependence of Eq.~(\ref{varene}) on $C$ and
$U$ through the optimization of $a_j$, i.e., the local density
modulation depends on $C$ and $U$, which yields the nontrivial mixing
properties.
This effect therefore cannot be described by the tight-binding model,
which has been used in previous studies on a two-component BEC in
an optical lattice~\cite{Jin, Roscilde, Ruo, Buonsante, Maska,
  Shrestha, Buonsante2, Wang, Ozaki, Suthar}.

From the energy curves in Fig.~\ref{fig:fig1}(c), the energetic
stability of the state against phase separation can be understood in a
diagrammatic manner~\cite{Porter}.
Let us consider a point on an energy curve $(C_0, \varepsilon(C_0))$,
and consider a situation in which the entire system is occupied by this
state.
Although this state is alternately modulated in the $x$ direction, as
shown in Fig.~\ref{fig:fig1}(d), with coarse-graining on a scale much
larger than the modulation wavelength, the two components are
uniformly mixed on average, and we refer to this state as a ``globally
mixed state''.
Suppose that this phase with $C_0$ separates into two phases with
$C_+ (> C_0)$ and $C_- (< C_0)$.
It can be shown that the energy of the separated state is given by
$\varepsilon_{\rm sep} = [\varepsilon(C_-) (C_+ - C_0) +
  \varepsilon(C_+) (C_0 - C_-)] / (C_+ - C_-)$ (See Supplemental
Material), which corresponds to the intersection point between the
vertical line $C = C_0$ and the line connecting the two points
$(C_+, \varepsilon(C_+))$ and $(C_-, \varepsilon(C_-))$.
When this energy $\varepsilon_{\rm sep}$ is larger (smaller) than
$\varepsilon(C_0)$, the globally mixed state with $C_0$ is
energetically stable (unstable) against phase separation into two
phases with $C_\pm$.
Thus, within the region of $\partial^2 \varepsilon / \partial C^2 <
0$, the globally mixed state is always unstable against phase
separation.
If $\partial^2 \varepsilon / \partial C^2 > 0$ and there exist $C_\pm$
such that $\varepsilon_{\rm sep} < \varepsilon(C_0)$, then the
globally mixed state is metastable.

For $U = 1.2$, the energy curve with respect to $C$ acquires a
concave-convex shape, as shown by energy curve IV in
Fig.~\ref{fig:fig1}(c).
In this case, the globally mixed state for, e.g., $C = 0.1$ is
unstable against phase separation.
From the above consideration, the most stable (lowest-energy)
separated pair of phases is given by the tangential line shown in
Fig.~\ref{fig:fig1}(c), which gives $C_- = 0$ and $C_+ \simeq 0.426$
(two circles on the line).
This indicates that if the globally mixed state with $C = 0.1$ is
prepared, it separates into two phases; one phase is occupied by only
component 2 ($C_- = 0$) and the other phase is occupied by both
components ($C_+ \simeq 0.426$).
This separated state is referred to as a mixed-bubble state, which was
first predicted in a system with beyond-mean-field effects~\cite{15}.
Here, the mixed-bubble state emerged even in a dilute system, in which
simple mean-field theory is applicable.

To confirm the results of the variational analysis, we numerically
solve the coupled Gross-Pitaevskii (GP) equation,
\begin{equation}
i \frac{\partial\Psi_{j}}{\partial t} = \left(
-\frac{\nabla^2}{2} + V_{j} + g \vert\Psi_{j}\vert^{2}
+ g_{12}\vert \Psi_{j'}\vert^{2} \right) \Psi_{j},
\label{GP}
\end{equation}
where $(j, j') = (1, 2)$ and $(2, 1)$.
Note that the results in Fig.~\ref{fig:fig1} are not dependent on the
dimensionality, and here we consider a two-dimensional system.
The system is discretized into a mesh with $dx = dy = 2\pi / 64$ and
the time step is typically $dt=0.001$.
The GP equation is integrated using the pseudospectral
method~\cite{28} with the periodic boundary condition.

\begin{figure}[tb]
\includegraphics[width=7.0cm]{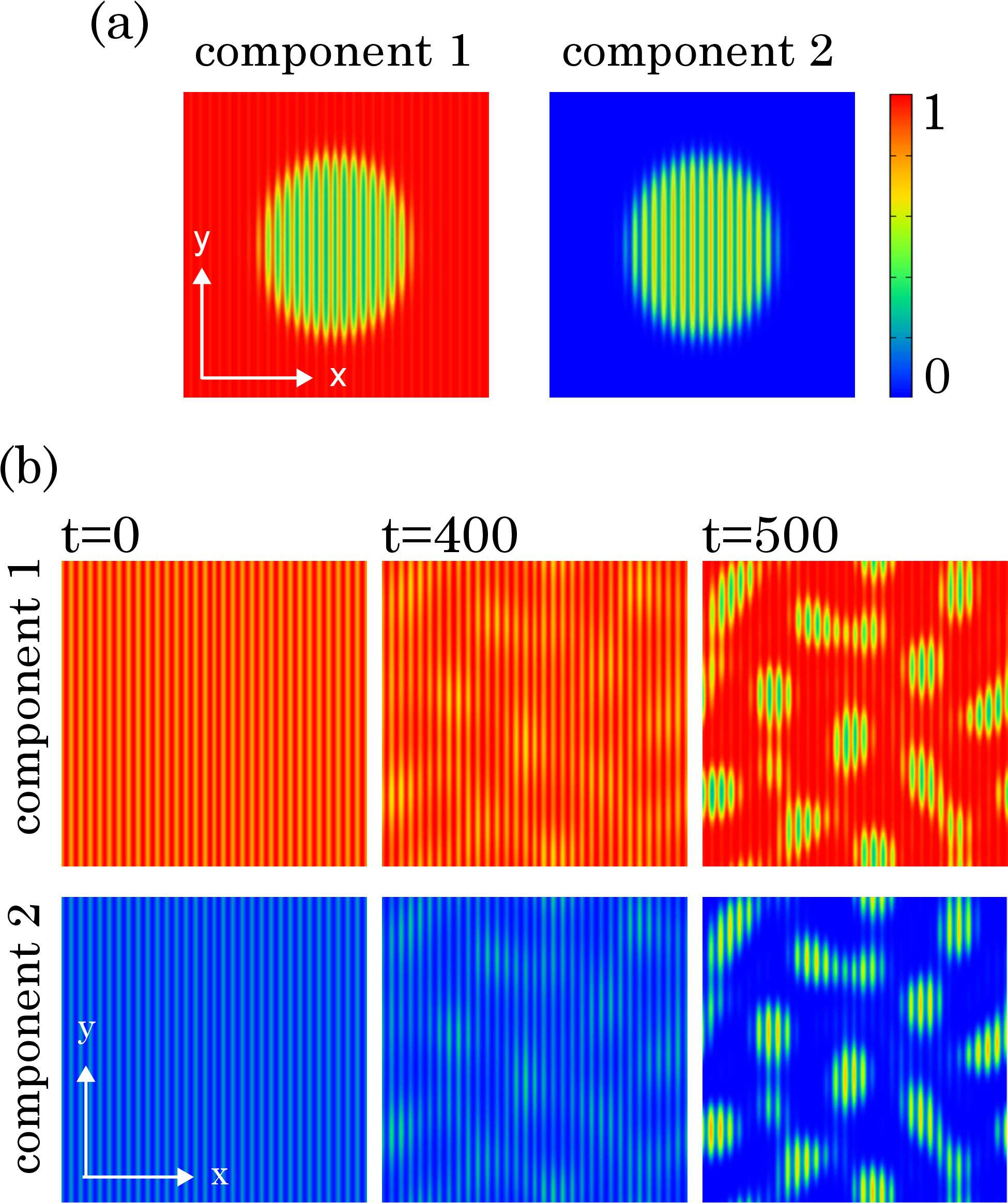} 
\caption{
  Mixed-bubble state for $C = 0.1$, $g = 15$, $g_{12} = 15.3$, and $U
  = 1.2$, which corresponds to the energy curve IV in
  Fig.~\ref{fig:fig1}(c).
  (a) Ground state obtained by solving the GP equation.
  (b) Dynamics of the mixed-bubble formation.
  The initial state is the ground state for $g = g_{12} = 15$.
  At $t = 0$, $g_{12}$ is suddenly changed to 15.3. 
  The size of each panel is $(32\pi)^2$.
\label{fig:bubble}}
\end{figure}

First, we solve the imaginary-time evolution of Eq.~(\ref{GP}) to
obtain the ground state, in which $i$ on the left-hand side is
replaced by $-1$.
Figure~\ref{fig:bubble}(a) shows the density distributions,
$|\Psi_1|^2$ and $|\Psi_2|^2$, of the ground state for $U = 1.2$ and
$C = 0.1$.
As expected from the variational results, the ground state is the
mixed-bubble state, which contains a single bubble of a mixed phase
surrounded by component 2.
The bubble in Fig.~\ref{fig:bubble}(a) is slightly elongated in the
$y$ direction, which implies that the interfacial tension between the
two phases is anisotropic due to the periodic potential in the $x$
direction.
Figure~\ref{fig:bubble}(b) shows the real-time evolution, in which the
initial state is the ground state for $g = g_{12} = 15$ and $C = 0.1$
with small random noise.
This state is a globally mixed state, as shown in the leftmost panels
of Fig.~\ref{fig:bubble}(b).
At $t = 0$, $g_{12}$ is suddenly changed to 15.3 (the same condition
as in Fig.~\ref{fig:bubble}(a)), and the mixed bubbles are dynamically
formed by spinodal decomposition.

\begin{figure}[tb]
\includegraphics[width=8.0cm]{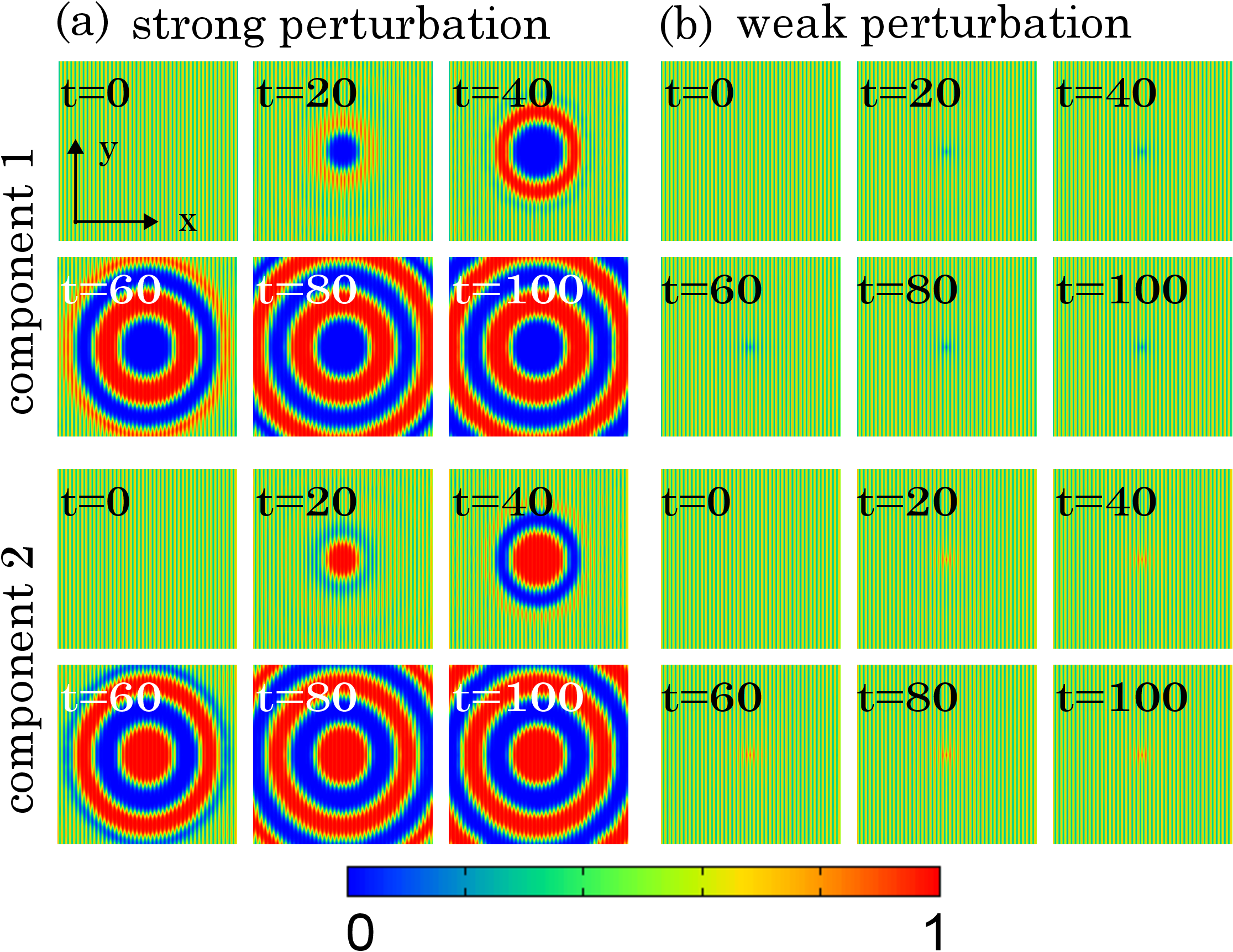}
\caption{
  Dynamics for $C = 0.5$, $g = 15$, $g_{12} = 15.3$, and $U = 1$.
  The energy curve for these parameters is shown in
  Fig.~\ref{fig:fig1}(c) (curve II).
  The initial state is the metastable state with $C = 0.5$ (circle in
  Fig.~\ref{fig:fig1}(c)).
  At $t = 0$, a local perturbation potential $A e^{-(x^2 + y^2)}$ is
  added with (a) $A = 1$ and (b) $A = 0.05$.
  The size of each panel is $(64\pi)^2$ with the origin at the
  center.
\label{fig:ncl}}
\end{figure}

For $U = 1$, the energy curve acquires the shape shown in
Fig.~\ref{fig:fig1}(c) (curve II).
Around $C = 0.5$, the energy curve is convex and therefore the
globally mixed state is stable against a small change in $C$ ($|C_\pm
- C_0| \ll 1$).
However, the true ground state is the totally separated phases with $C
= 0$ and $C = 1$, and the globally mixed state with $C \simeq 0.5$ is
a metastable state.
Figures~\ref{fig:ncl}(a) and \ref{fig:ncl}(b) show the dynamics
that start from the metastable state with $C = 0.5$, where a local
perturbation potential $A e^{-(x^2 + y^2)}$ is added at $t = 0$ to
trigger the nucleation.
Figure~\ref{fig:ncl}(a) shows the case of $A = 1$;
the phase separation is triggered around the center by the
perturbation potential, and the concentric phase separation extends 
outward, which is the phase separation by nucleation.
In the case of a smaller perturbation ($A = 0.05$), as shown in
Fig.~\ref{fig:ncl}(b), the density distributions around the center are
slightly modified by the perturbation potential, which is insufficient
to trigger the phase separation.
This corroborates the existence of an energy barrier against
nucleation.

\begin{figure}[tb]
\includegraphics[width=7.0cm]{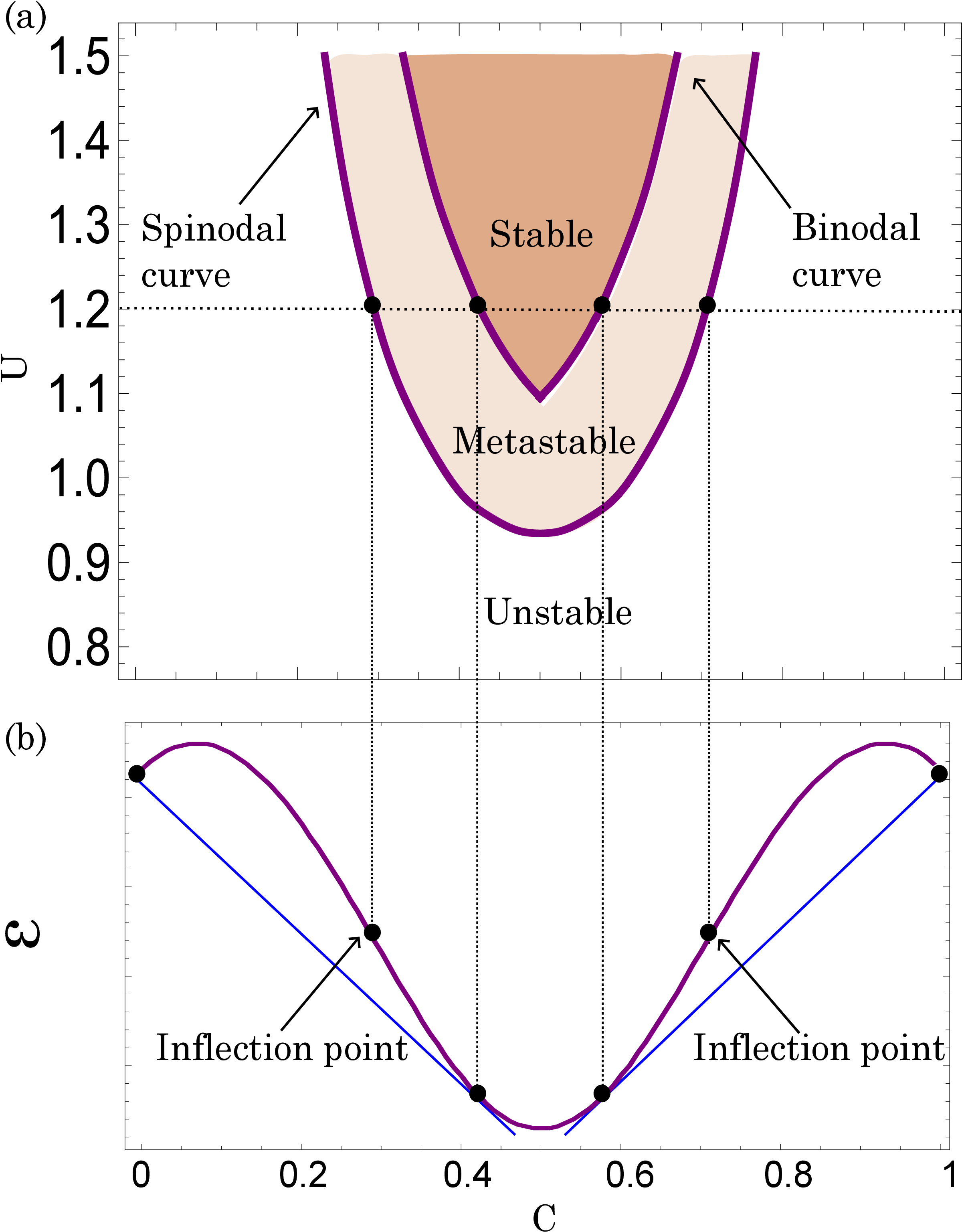} 
\caption{
  (a) Stability of the globally mixed state with the composition $C$
  for the potential strength $U$, $g = 15$, and $g_{12} = 15.3$.
  The spinodal (binodal) curve divides the unstable (stable) and
  metastable regions.
  The parameters used in Figs.~\ref{fig:bubble} and \ref{fig:ncl} are
  marked by the open circles.
  (b) Energy curve $\varepsilon(C)$ for $U = 1.2$.
  The inflection points (filled circles) trace the spinodal curve in
  (a).
  The tangent points (squares and triangles) for the solid lines give
  the mixed bubble states.
  The points marked by the triangles trace the binodal curve in (a).
\label{fig:pd}}
\end{figure} 

Figure~\ref{fig:pd}(a) depicts the stability of the globally mixed
state with composition $C$ for the potential strength $U$.
For the region in which the energy curve $\varepsilon(C)$ is concave,
$\partial^2 \varepsilon / \partial C^2 < 0$, the globally mixed state
is unstable against phase separation, and the inflection points
(circles in Fig.~\ref{fig:pd}(b) for $U = 1.2$) trace the spinodal
curve in Fig.~\ref{fig:pd}(a).
For $U \lesssim 0.94$, $\varepsilon(C)$ is concave everywhere and
there are no inflection points.
The tangential lines, as shown in Fig.~\ref{fig:pd}(b), give the
mixed-bubble states, and a globally mixed state for $C$ between the
square and triangle in Fig.~\ref{fig:pd}(b) has higher energy than the
mixed-bubble state.
Therefore, the region between the circle and triangle is metastable.
For $U \simeq 1.1$, the states with $C = 0$, $C = 0.5$, and $C = 1$
become degenerate (curve III in Fig.~\ref{fig:fig1}(c)), and these
three phases (occupied only by component 0 or 1, or equally mixed) can
coexist.
 
Experimentally, the phenomena presented here can be realized by a
two-component BEC with scattering lengths that satisfy the immiscible
condition $a_{12}^2 - a_{11} a_{22} > 0$.
(See Supplemental Material for an example of an experimental system.)
A box-like potential rather than a harmonic potential is suitable to
avoid the complexity that arises from inhomogeneous density.
A quasi-two-dimensional (or one-dimensional) system is convenient to
observe the spatial density pattern, and also to suppress the total
number of atoms; however, the dimensionality is not crucial for the
present phenomena.

In conclusion, we have proposed a method to control the mixing
properties of two fluids.
The mixing energy can be changed by modulating the densities on a
small scale using component-dependent external potentials, which
alters the global mixing properties.
This method was applied to a two-component BEC of dilute gases.
Although this system originally has a trivial mixing property, the
energy curve acquires a concave-convex shape with respect to the
composition $C$ by the present method, and spinodal and binodal
physics emerge.
As a result, the mixed-bubble state (Fig.~\ref{fig:bubble}), which has
only been predicted for a system with large quantum fluctuation,
becomes possible for a simple dilute system.
The modification of the energy curve also results in a metastable
state that undergoes phase separation via nucleation due to a finite
local perturbation (Fig.~\ref{fig:ncl}).
The present method is not restricted to quantum fluids and may also
be applied to classical immiscible fluids, such as oil and water.

\begin{acknowledgments}
This work was supported by JSPS KAKENHI Grant No. JP23K03276.
\end{acknowledgments}

\end{document}